\documentclass[aps,prl,twocolumn,letterpaper,superscriptaddress,showpacs]{revtex4}
\usepackage{amsmath, amsthm}
\usepackage{graphicx,latexsym}
\usepackage{dcolumn}
\usepackage{color}

\begin{document}

\title{Lifshitz Transition and Superconductivity Enhancement in High Pressure \textit{cI}16 Li}

\author{Chia-Hui Lin}
\affiliation{Condensed Matter Physics and Materials Science Department,
Brookhaven National Laboratory, Upton, New York 11973, USA}
\affiliation{Department of Physics and Astronomy, Stony Brook University, Stony Brook, New York 11794, USA}

\author{Tom Berlijn}
\affiliation{Condensed Matter Physics and Materials Science Department,
Brookhaven National Laboratory, Upton, New York 11973, USA}
\affiliation{Center for Nanophase Materials Sciences and Computer Science and Mathematics Division, Oak Ridge National Laboratory, Oak Ridge, TN 37831-6494}

\author{Wei Ku}
\affiliation{Condensed Matter Physics and Materials Science Department,
Brookhaven National Laboratory, Upton, New York 11973, USA}
\affiliation{Department of Physics and Astronomy, Stony Brook University, Stony Brook, New York 11794, USA}
\date{\today}

\begin{abstract}
The Fermi surface topology of \textit{cI}16 Li at high pressures is studied using a recently developed first-principles unfolding method.
We find the occurrence of a Lifshitz transition at $\sim$43 GPa, which explains the experimentally observed anomalous onset of the superconductivity enhancement toward lowered pressure. Furthermore we identify, in comparison with previous reports, additional nesting vectors that contribute to the \textit{cI}16 structural stability. Our study highlights the importance of three-dimensional unfolding analyses for first-principles studies of Fermi surface topologies and instabilities in general.
\end{abstract}

\pacs{71.15.-m, 71.18.+y, 71.20.Dg, 74.62.Fj}
\maketitle

\begin{figure}
    \centering
  \includegraphics[width=1\columnwidth,clip=true]{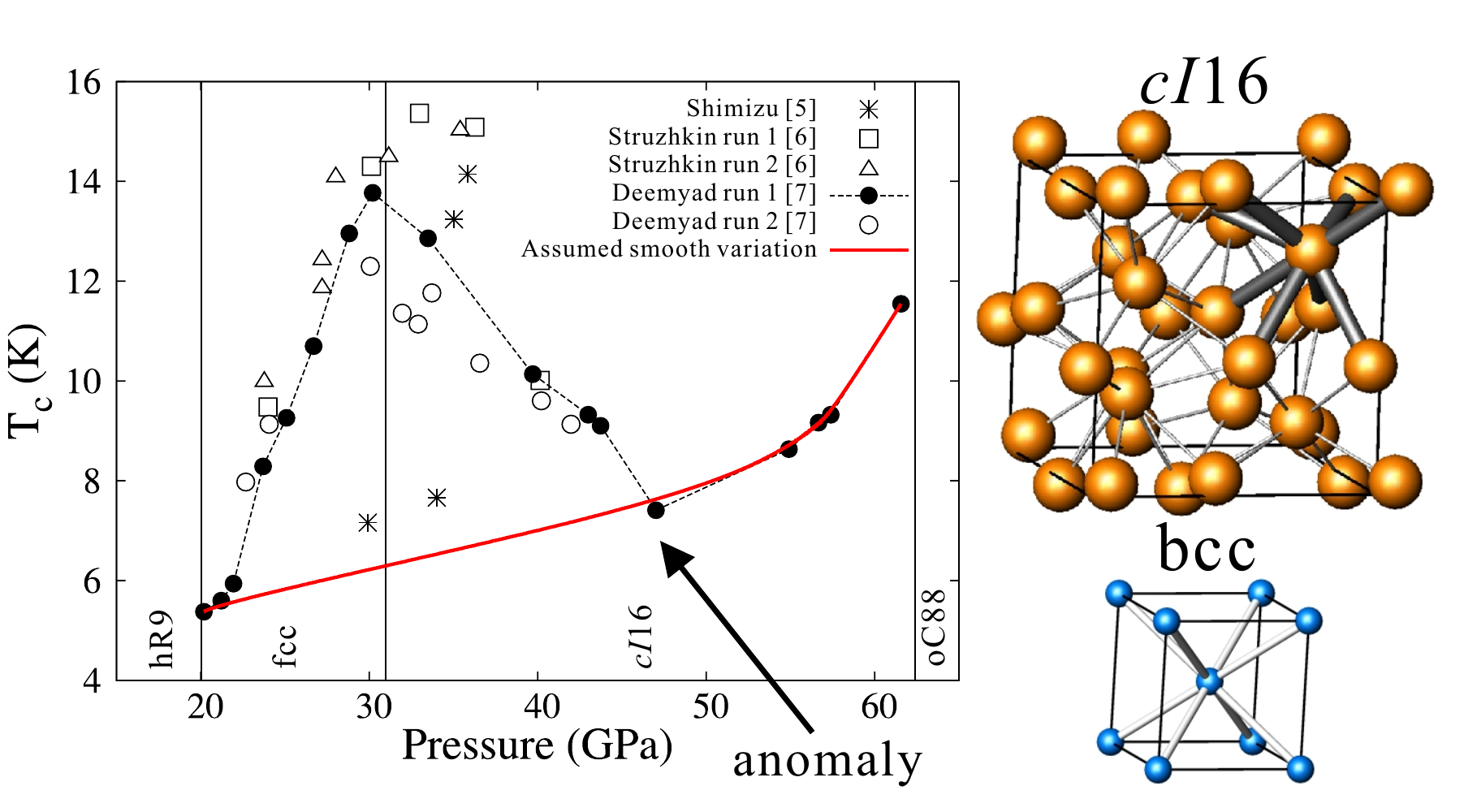}
  \caption{
(Left) Experimental phase diagram of Li with the vertical structural transition lines from Refs.~\onlinecite{Shimizu:0597, Struzhkin:1213, Deemyad:7001, Guillaume2011,Hanfland:0174}. The assumed smooth variation (solid red curve) is simulated by the basis spline method on the lowest two and highest five pressure data points. (Right) Conventional \textit{cI}16 (top) and bcc (bottom) unit cells.
}
  \label{fig1}
\end{figure}

In 2007, almost one century after Onnes' first discovery of superconductivity \cite{Onnes1911}, Li eventually joined ambient-pressure superconductors with a critical temperature ($T_c$) down to 0.4 mK \cite{Tuoriniemi:0187}. Its high-pressure $T_c$ up to 16 K surprisingly tops all elements at similar pressures.
Being an elemental conventional superconductor with a rich phase diagram (c.f. Ref. \onlinecite{Guillaume2011} and Fig.~1), Li is a natural touchstone for first-principles methods of determining $T_c$. Certain breakthroughs have been established in the frameworks of Eliashberg theory  \cite{Christensen:1861,Iyakutti:2504,Maheswari:3227,Kasinathan2006,Shi:4516,Yao:4524,Bazhirov2010,Bazhirov2011} and superconducting density functional theory \cite{Profeta:7003,Akashi2013} to satisfactory agreement with the experimental trend in the correct order of magnitude.
Such phonon-mediated superconductivity can be strongly tied with lattice instabilities. For both ends of the collected data in Fig.~\ref{fig1}, the disappearance of superconductivity at 20 and 62 GPa can undoubtedly be attributed to the occurrence of structural transitions. At the unusual $T_c$ maximum around 30 GPa a substantial phonon softening maximizes the electron-phonon coupling \cite{Bazhirov2010} and causes a transition to the \textit{cI}16 structure. Nonetheless, it remains puzzling that the $T_c$ subsequently plunges into a minimum between 44 and 47 GPa in the absence of structural and electronic transitions or singular phonon behavior \cite{Yao:4524}.

A peculiar change in the electronic structure is believed to account for this anomalous pressure, below which the superconductivity is dominantly controlled by the fermiology. Between 20 and  30 GPa the superconductivity is enhanced by accumulating FS nesting in the fcc side \cite{Kasinathan2006,Bazhirov2010}, reaching a 16 K maximum. In the subsequent \textit{cI}16 phase $T_c$ decreases due to a monotonic FS depletion \cite{Yao:4524}.
The depletion mainly comes form the nature of the \textit{cI}16 charge density wave (CDW), which consists of a distortion along the cube diagonal within its unit cell composed of $2\times 2 \times 2$ regular body-centered cubic cells (see right panels in Fig.~\ref{fig1}). Above the anomalous pressure the FS depletion persists \cite{Yao:4524} opposite to the $T_c$ growth in the experiment. Consequently the major influence on superconductivity is taken over by another effect that is insensitive to the fermiology. Its impact on the $T_c$ (tentatively assumed to be smooth) is depicted by the red curve in Fig.\ref{fig1}.
The anomaly can be regarded as an onset pressure, below which the fermiology resumes dominance and promptly increases $T_c$. Therefore, the elaboration on the FS evolution around the pressure anomaly is highly demanded for resolving the missing puzzle piece in the phase diagram.

In this letter, we identify the occurrence of a Lifshitz transition (LT) in high-pressure \textit{cI}16 Li from first-principles.
Our unfolded Fermi surface shows that when the pressure is lowered across the anomalous point at $\sim$43 GPa, there appears new FS pockets.
This gives rise to a non-analytic superconductivity enhancement, which we verify numerically and find to be in quantitative agreement with experimental observations.
Our results also visually decode the effects of three-dimensional charge density waves on the Fermi surface.
We find the \textit{cI}16 nesting vectors [200]$\frac{2\pi}{a_{cI16}}$ and $[110]\frac{2\pi}{a_{cI16}}$ to be as relevant as the previously reported $[211]\frac{2\pi}{a_{cI16}}$~\cite{Prieto:0021}.
These intriguing and direct observations demonstrate the general value of Fermi surface unfolding in the study of topological features associated with symmetry breaking phase transitions in materials.

\begin{figure}
    \centering
   \includegraphics[width=1.0\columnwidth,clip=true]{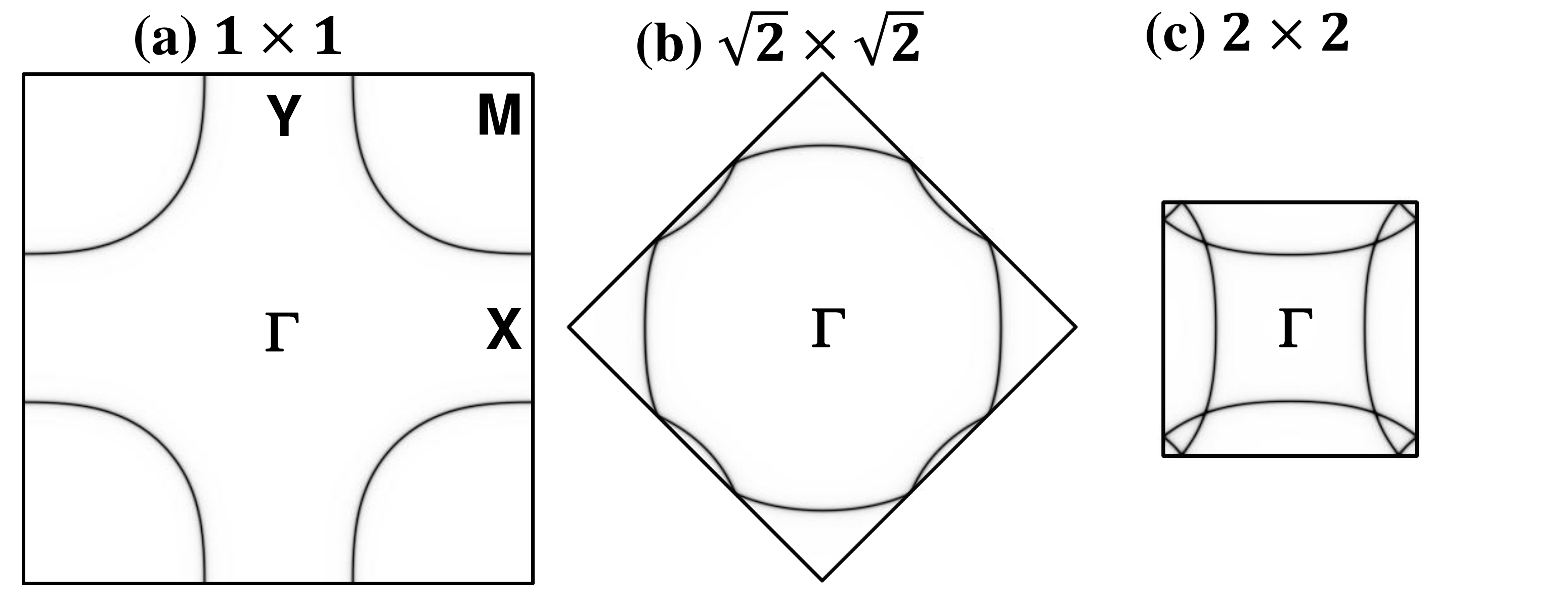}
  \caption{Illustration of Fermi surface folding issue. (a) Fermi surface of half-filling two-dimensional one-band tight-binding model with hopping integrals $t^{\prime}/t=-0.25$ and $t^{\prime \prime}/t = 0.125$. (b) the same model solved with a $\sqrt{2}\times \sqrt{2}$ supercells. (c)
  the same model solved with a $2\times 2$ supercells. Recovery from (b)\&(c) to (a) can be achieved by FS unfolding.}
  \label{fig2}
\end{figure}

A commonly encountered difficulty in the study of Fermiology of large unit cell systems is the hinderance from severe band folding. For example in Fig.~\ref{fig2} we illustrate with a two-dimensional toy model the complexity of FS folding that arises from the arbitrary choice of lattice periodicity even in the absence of any physical broken symmetry. This reflects the fact that Bloch wavefunctions take more information out of the folded FS as the supercell size grows.
This difficulty can be overcome by the recently developed unfolding method \cite{Ku2010}, which is aimed to restore the informative Green's function in a normal cell basis.
Thus the one-particle spectral function in the eigenstates basis, $A_{KJ,KJ}(\omega)$, can be converted into a reference basis $A(k, \omega) = \sum_{KJn}|\langle kn| KJ\rangle|^2 A_{KJ,KJ}(\omega)$.
Here, $K$/$k$ is the crystal momentum of the original/reference system, $J$ the eigenstate band index, and $n$ the Wannier orbital index. With the use of Wannier functions the spectral weight magnitude $|\langle kn| KJ\rangle|^2$ reduces to a simple structure factor that is readily evaluated. \cite{Ku2010}

\begin{figure}
    \centering
   \includegraphics[width=1.0\columnwidth,clip=true]{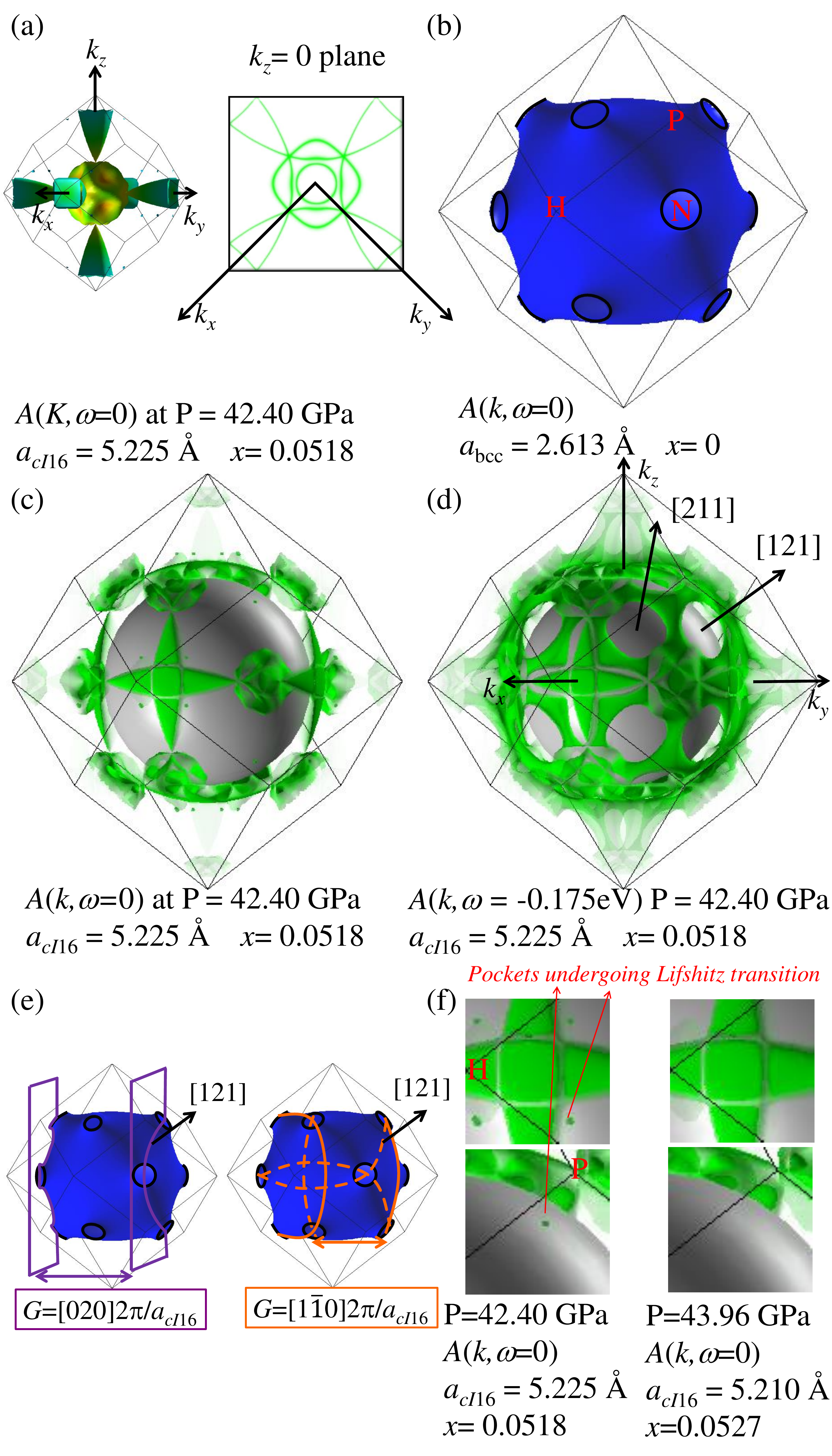}
  \caption{
Fermi surfaces of (a) \textit{cI}16 Li at 42.40 GPa and (b) its reference CDW-free bcc Li.
(c) Unfolded \textit{cI}16 Fermi surface and (d) energy isosurface at -0.175 eV.
The opacity at each $k$ point represents the spectral weight with the most/least transparent spots corresponding to 0.02/0.88.
For the purpose of clarity, the grey balls in (c-d) are utilized to block image from the back side.
(e) Fermi surface intersections with the nesting planes spanned by [020]$\frac{2\pi}{a_{cI16}}$ (left) and [1$\bar{1}$0]$\frac{2\pi}{a_{cI16}}$ (right).
(f) Enlargements showing pocket appearance (i.e. Lifshitz transition).
}
  \label{fig3}
\end{figure}

Specifically in \textit{cI}16 Li, the unfolding method allows to reveal detailed information of the Fermi surface topology.
Density functional theory calculations are performed with the WIEN2K \cite{Wien2k} implementation of the full potential linearized augmented plane wave method in the local density approximation.
Then, the symmetry-respecting Wannier functions \cite{Ku2002} of Li $s$ and $p$ orbitals are constructed within Hilbert space between -5 to 30 eV. The lattice constant $a_{cI16}$ and atomic displacement $x$ ($\approx$ 0.05 and increasing with pressure) away from the bcc structure are obtained from the experimental values \cite{Hanfland:0174}. Then, to obtain the unfolded electronic structure in \textit{cI}16 Li, the natural choice for the reference basis is the bcc lattice as shown in the right of Fig. \ref{fig1}. In this study, we use opacity in the three-dimensional reciprocal space to depict the spectral weight magnitude on the unfolded FS and energy isosurfaces.

The \textit{cI}16 FS in Fig.~\ref{fig3}(a, left) gives a good example of an overwhelmingly folded FS in the \textit{cI}16 Brillouin zone (BZ) which is eight times smaller than that of the bcc lattice.
An additional complication of folding in three dimensions (compared to lower dimensions) is that when FS sheets are intertwined, they block each other. The cross sectional view on the $k_z=0$ plane in the right of Fig.~\ref{fig3}(a) shows that the outer FS sheet is blocking two inner FS sheets. In contrast, the unfolded FS in Fig.~\ref{fig3}(c) recovers the resemblance to Fig.~\ref{fig3}(b), the FS of the fictitious CDW-free bcc system with $a_{\mathrm{bcc}}=a_{cI16 }/2$. For example the necks at the high symmetry points N resembling the FS of copper \cite{Kasinathan2006, Prieto:0021} are still clearly recognizable in the unfolded FS. The grey balls in Fig.~\ref{fig3}(c)(d)(f) are used to block the image from the back side for the better visualization and do not containing any visible weight inside.

The effect of the CDW is clearly demonstrated in Fig.~\ref{fig3}(c).
The comparison between Fig.~\ref{fig3}(b) and (c) shows that the CDW-induced gaps substantially deplete the DOS around the Fermi energy for energy gain.
Moreover these gaps offer the opportunity to change the FS topology.
In the bcc reference basis, the CDW couples each $k$ point to eight $k+G$ points by the potential $V_{\mathrm{CDW}}^{G}$, where $G$ represents the eight reciprocal lattice vectors of \textit{cI}16 unit cell that lie in the bcc BZ. In the presence of substantial hybridization, spectral weight is re-distributed between the eight coupled $k$ points and appears as transparent sheets of Fermi surface. The opacity of these sheets reflects the strength of $V_{\mathrm{CDW}}^{G}$. Therefore, the unfolded FS provides a systematic way to detail its coupling with the order parameter of the broken translational symmetry.

Due to the serious FS depletion, the identification of the relevant CDW wave vector $G$'s is made easier if we focus on the unfolded energy isosurface of 0.175 eV below the Fermi energy in Fig.~3(d).
The effects of $V_{\mathrm{CDW}}^{G}$ can be understood from considering the intersections of the nesting planes $\pm\frac{1}{2}G$ with the FS of the fictitious CDW-free bcc system as shown in Fig.~\ref{fig3}(e). The Bloch states at these intersections are degenerate and therefore will be gapped out most strongly. For example, the left of Fig.~\ref{fig3}(e) shows the degenerate Bloch states that are nested by $[020]\frac{2\pi}{a_{cI16}}$ (representing the $G's$ symmetry-related to [200]$\frac{2\pi}{a_{cI16}}$) and explains the cut traces connecting any two neighboring $N$ points in Fig.~\ref{fig3}(d). Another CDW component is illustrated by the pair of solid/dashed orange planes spanned by $G=[1\bar{1}0]\frac{2\pi}{a_{cI16}}$ (representing the $G's$ symmetry-related to [110] $\frac{2\pi}{a_{cI16}}$) in the right of Fig.~\ref{fig3}(e). If we collect multiple symmetry-related cut marks, $V_{\mathrm{CDW}}^{[110]}$ is concluded to be responsible for the \textit{cross-like} mark on every $H$ point in Fig.~3(d).

The most prominent FS depletion takes place on the 24 large holes along the [211] direction had been previously been assigned to the $G=[211] \frac{2\pi}{a_{cI16}}$ nesting vector \cite{Prieto:0021}. Indeed $V_{\mathrm{CDW}}^{211}$ partially contributes to those substantial gaps.
However, both $V_{\mathrm{CDW}}^{200}$ and $V_{\mathrm{CDW}}^{110}$ also conspire to cause DOS depletion because their cut marks pass those holes too,
as exemplified with the arrows in the [121] direction in Fig.~\ref{fig3}(e). Contrary to low-dimensional physics, the FS nesting importance is diluted by the consideration of the phase space, which takes into account all relevant $V_{\mathrm{CDW}}^{G}$'s. Fig.~3(d) indicates that the phase space affected by $V_{\mathrm{CDW}}^{200}$ and $V_{\mathrm{CDW}}^{110}$ is at least comparable with the previously emphasized $V_{\mathrm{CDW}}^{211}$, and so is the energy gain.

\begin{figure}
    \centering
   \includegraphics[width=1\columnwidth,clip=true]{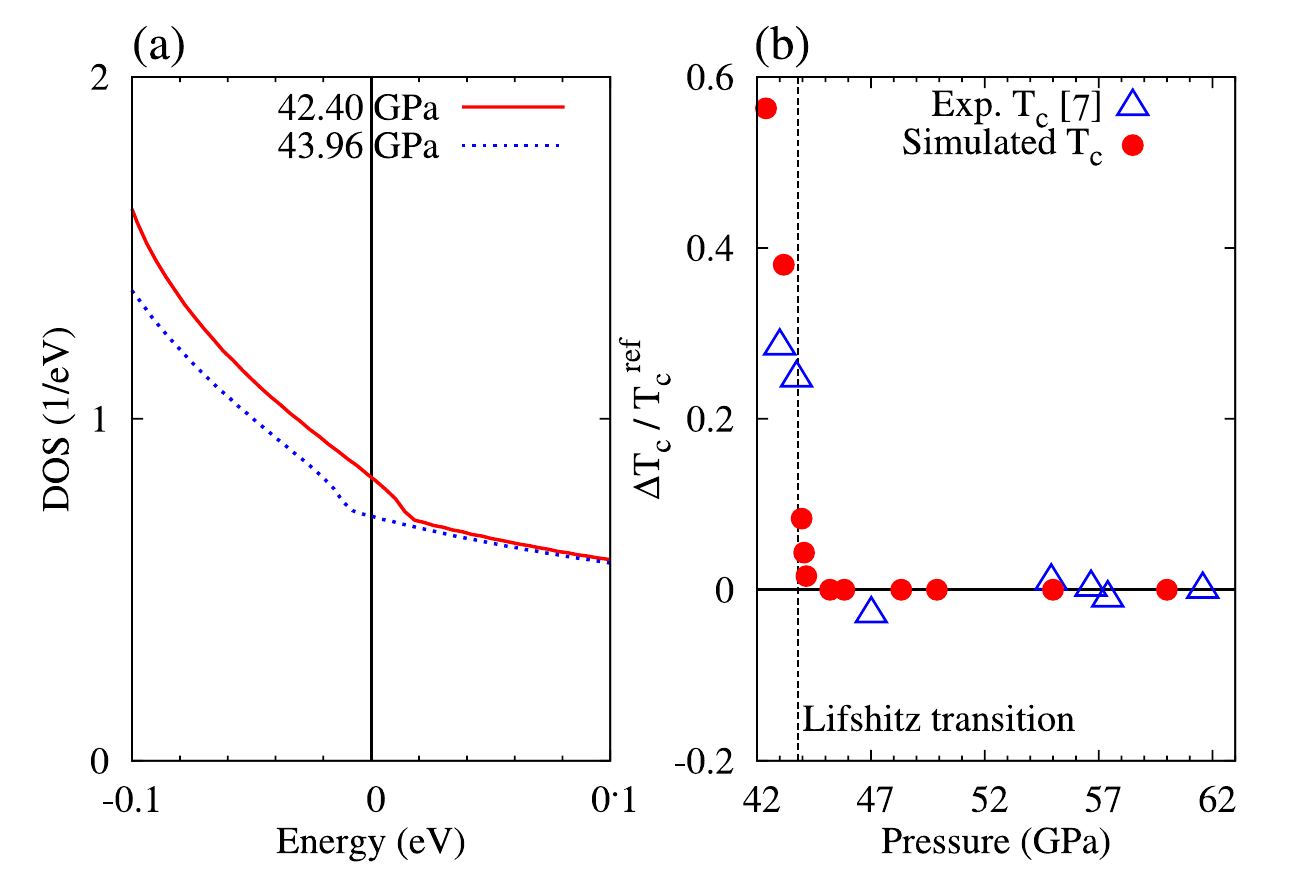}
  \caption{
(a) DOS at 42.40 and 43.96 GPa.
(b) Experimental ~\cite{Deemyad:7001} and simulated $\Delta T_c/T^{ref}_c$ versus pressure for \textit{cI}16 Li.
  \label{fig4}
  }
\end{figure}

Next we investigate the FS evolution of \textit{cI}16 Li near the anomalous pressure in the phase diagram shown in Fig.~\ref{fig1}.
Interestingly, we find the emergence of new FS pockets around the N and P at the pressure between 42.40 and 43.96 GPa in Fig.~\ref{fig3}(f).
This topological change of FS is termed Lifshitz transition \cite{Lifshitz:6401}.
Although they are seemingly small pockets, the LT is known to cause a dramatic change in physical observables within a small pressure window \cite{Chu:0214}.
It also has been proven to induce non-analytic behavior on the Fermi surface density of states (DOS) \cite{Lifshitz:6401} and conventional superconductivity \cite{Chu:0214,Makarov:1151}.
Recently, LTs have been applied to engineer the topological order in topological insulators Bi$_{1-x}$Sb$_{x}$ \cite{Hasan2010}.
Also, doping-induced LTs have been proposed to explain the diverging cyclotron mass in high $T_c$ cuprates \cite{Norman2010} and the vanishing transport anisotropy in the novel Fe-based superconductors \cite{Yi2009}.

The LT in \textit{cI}16 Li gives rise to an anomalous contribution in the DOS, $\delta g(\omega)\propto \sqrt{E_c-\omega}$ as $\omega \rightarrow E_c$ \cite{Lifshitz:6401}, originating from the extra pockets.
Here the pressure-dependent $E_c$ is the critical energy at which the pockets disappear. The result in Fig.~\ref{fig4}(a) not only confirms the correct square root behavior but also is consistent with the observation in Fig.~\ref{fig3}(f) that the Fermi energy $E_F$ is located above (below) $E_c$ at 43.96 (42.40) GPa. It is important to stress that the occurrence of the LT is confined to an infinitesimally small part of the $k$ space. Therefore to capture the singular behavior of the DOS numerically a $400\times400\times400$ k mesh has been employed. Such an extremely fine k-mesh is only possible by the use of Wannier interpolation and is otherwise inaccessible in standard first-principles calculations.

In absence of any symmetry breaking, this LT is the natural explanation of the anomalous onset (between 44 and 47 GPa) toward lower pressures.
Below this pressure the $T_c$ displays a sharp upturn followed by a continuous enhancement (c.f. Fig.~\ref{fig1}) due to the extra available DOS at the Fermi energy.
Makarov and Baryakhtar have shown that in the weak coupling Bardeen-Cooper-Schrieffer (BCS) theory the anomalous part of the DOS, $\delta g(\omega)$, leads to an asymmetric variation of the $T_c$. \cite{Makarov:1151}
This physical picture can be applied to \textit{cI}16 Li.
At pressures as high as 60 GPa, $E_c$ is far below $E_F$ and has absolutely no effect on superconductivity.
But at a critical pressure above the LT, $E_F$ will fall right below $E_c+\omega_D$ with $\omega_D$ the Debye frequency. At this pressure $T_c$ will suddenly start to surge because the extra DOS $\delta g(\omega)$ impacts the $T_c$ exponentially. As the pressure is further lowered, the small pockets grow and contribute to the $T_c$ enhancement in a continuous manner.

The physical picture described above can be formulated into a quantitative agreement in the superconductivity enhancement between the experimental data and our simulation.
For the purpose of demonstration, we tentatively separate the $\delta g(\omega)$ correction on the superconductivity as $\Delta T_c$ from the other smooth pressure-dependent part $T_c^{ref}$. For the experimental part the smoothly varying $T_c^{ref}$ is attributed to the solid red curve in Fig.~\ref{fig1}, and $\Delta T_c$ is defined as its difference from the total $T_c$. For the theoretical part we can simulate $T_c$ and $T_c^{ref}$ by numerically solving the BCS gap equation with and without the non-analytic contribution $\delta g(\omega)$ respectively and obtain $\Delta T_c$ as their difference.
In order to simulate the DOS in more pressure conditions, the lattice constant and atomic displacement are refined by linear fitting with respect to the experimental pressures in Ref.~\onlinecite{Hanfland:0174}. The effective pairing potential $V_{ee}$ in the gap equation ( $1 = V_{ee} \sum_{\omega=E_F-\omega_D}^{E_F+\omega_D} \frac{g(\omega)}{2 \omega} \tanh \frac{\omega}{2T_c}$) is fixed to be 345 meV to obtain $T_c\sim 9$ K at the LT.
The Debye frequency $\omega_D=21.5$ meV is chosen to match the characteristic phonon frequency in Ref.~\onlinecite{Yao:4524}.
The resulting $\Delta T_c/T_c^{ref}$ ratio demonstrates excellent agreement in Fig.~\ref{fig4}(b) as the sudden rise right below the LT pressure is well matched.
Therefore, the puzzling strong enhancement of superconductivity below the anomalous pressure can now be understood as a consequence of the LT.

In summary, we identify a Lifshitz transition  at $\sim$43 GPa in \textit{cI}16 Li to be responsible for the experimentally observed onset pressure, below which the superconductivity is highly enhanced.
This is achieved via the unfolding method, which significantly facilitates the Fermi surface visualization. The implementation of three-dimensional Fermi surface unfolding shows the capability to decode the Fermi surface topology of \textit{cI}16 Li. Furthermore it allows to identify $V_{\mathrm{CDW}}^{200}$ and $V_{\mathrm{CDW}}^{110}$ in addition to the previously reported $V_{\mathrm{CDW}}^{211}$ as the important CDW components that stabilize \textit{cI}16 Li.
The three-dimensional unfolding of Fermi surfaces is expected to expedite a wide range of studies about topological evolution and structural stability of materials with broken translational symmetry in general.

\begin{acknowledgments}
We gratefully acknowledge the stimulating discussions with Olga Degtyareva.
This work is supported by the U.S. Department of Energy under contract DE-AC02-98CH10886.
TB was supported by DOE CMCSN and as a Wigner Fellow at the Oak Ridge National Laboratory.
\end{acknowledgments}

\end{document}